\documentstyle[epsfig,preprint,aps]{revtex}  
\tightenlines
\def\be{\begin{equation}}
\def\ee{\end{equation}}
\def\bea{\begin{eqnarray}}
\def\eea{\end{eqnarray}}
\def\beq{\begin{equation}}
\def\eeq{\end{equation}}
\begin{document}


\title{$R_{out}/R_{sid}$ and Opacity at RHIC \\[2ex]}
\author{Larry McLerran$^1$ and Sandra S. Padula$^{1,2}$ \\[2ex]}
%
%
\address{$^1$ 
Nuclear Theory Group, Brookhaven National Laboratory, Upton, NY 11973, USA \\
$^2$	Instituto de F\'{\i}sica Te\'orica, Universidade Estadual 
Paulista, S\~ao Paulo - SP, Brazil; \\
Theoretical Physics, FERMILAB, Batavia, IL 60510, USA}
\maketitle
\vskip-7.2cm\hfill
\begin{flushright}
BNL-69181\\
FERMILAB-Pub-02/077-T\\
IFT-P.029/02\\
\end{flushright}
\vskip 4.2cm
\begin{abstract}
One of the most dramatic results from the first RHIC run 
are the STAR results for $\pi^\pm \pi^\pm $ interferometry. They 
showed that the ratio of the so-called $R_{out}$ and $R_{sid}$ radii 
seem to decrease below unity for increasing transverse momentum of the 
pair ($K_T$). This was subsequently confirmed by PHENIX 
which also extended the $K_T$ range of the measurements. 
We consider here the effects of opacity of the nuclei on this
ratio, and find that such a small value is consistent with
surface emission from an opaque source.  

\end{abstract}

\section{Introduction}

	The striking  result on  $\pi^\pm \pi^\pm $ interferometry 
data obtained by STAR Collaboration\cite{star} on the ratio 
$R_{out}/R_{sid}$ is not yet understood.  This is the ratio of the 
{\sl outward} radius, measured in the correlation along the direction of the
sum of the particles' momenta, divided by the {\sl sideward} radius, 
reflecting the correlation perpendicular to the beam and perpendicular to 
the outward direction.  This result was subsequently
confirmed by PHENIX\cite{phenix}.  In an attempt to understand this 
result we consider in detail a generalization of the model used by 
Heiselberg and Vischer\cite{henn} to include effects of source opacity 
at CERN energies, later followed by Heinz and Tom\'a\u sik  
\cite{tomheiz1}.

The origin of the problem can be understood from the simplest model
of HBT\cite{hbt,heinzjacak}.  In this model we let particles be emitted from
the entire volume of the system.  In this case, the spatial size
which is probed by $R_{out}$ and $R_{sid}$ is of the same order of magnitude,
the radius of the system.  However, $R_{out}$ also has a spatial 
correlation built in on account of the time differences of different 
emissions\cite{phpb}.  
$R_{out}$ is the size scale measured along the particle's direction of motion.
Particles
emitted at much different times end up spatially separated from one another
in this direction.
We therefore expect that since the time separations are of
the order of the size of the system (which for practical purposes 
seems to be a reasonable approximation for order of magnitude estimates 
at RHIC), then $R_{out}/ R_{sid} > 1$.  Most simple models incorporating the above 
physics give the ratio to be about $R_{out}/R_{sid} \sim 2$.

The assumption that the particles are emitted from the entire volume of the
system is probably a bad one.  Data on $p_T$ distributions of particles
in STAR\cite{star} and PHENIX\cite{phenix} suggest that the system
is opaque to particles up to large transverse momenta.   
A more reasonable description might be blackbody surface emission.

To implement opacity, we consider a model where the matter is emitting
from a surface at a fixed radius.  The system is allowed to undergo
$1+1$ dimensional longitudinal expansion.  The details of this model 
and  its results are described in the fourth section. Many of the features
of the model we propose are embodied in the hydrodynamic computations of
Heinz and Kolb.\cite{upkolb}  The essential difference lies in 
the treatment of surface emission.

An important ingredient of this model is decoupling, which we
assume occurs at a well defined temperature.  In the surface emission 
model we use, this requires that, at the end of the process, the 
decoupling to occur throughout the transverse
volume of the system at a well defined time.  It turns out that, although 
it is a relevant ingredient (i.e., it contributes to the {\it averaging} 
over different emission radii), this is 
not so important in our computations because the surface emission
dominates.

In the second section, before turning to explicit model
computations, we review the method of the Covariant Current Ensemble 
formalism for computing two particle HBT correlation functions.
In this section, we try to be general enough to include the
physics needed for our various computations.

In the third section, we consider a simpler version of the above
model.  We compute the various radii for the case of a cylinder
emitting at a constant temperature $T$ for a finite time $\Delta t$, 
without introducing any radial flow.  
We will consider both the case of an opaque source and a transparent one.
In this oversimplified model, we show that in the opaque source limit,
that $R_{out} \sim  \sqrt{\Delta t^2 (K_T/E_K)^2 + (0.2 R_T)^2}$ 
and $R_{sid} \sim R_T$.
For the realistic case where $\Delta t \sim R_T$, $R_{out}/R_{sid} \sim 
\Delta t (K_T/E_K)/R_T$,
and this ratio can be less than one.  In the transparent limit
$R_{out} = \sqrt {\Delta t^2 (K_T/E_K)^2 + R_T^2 }$ and $R_{sid} \sim R_T$,  
this ratio is always larger than 1.

In the heavy ion experiments, the ratio $R_{out}/R_{sid}$ is larger than 1 
at AGS to SPS energies for $\pi^- \pi^-$ pairs\cite{heinzjacak,heinz}.
For  $\pi^+ \pi^+$ it is also larger than 1 
at AGS ($R_{out} / R_{sid} \approx 1.3$ - E859 Collab.\cite{heinzjacak,soltz})
but slightly smaller than 1 at SPS, according to NA44 data\cite{heinzjacak}.

Although the value of $R_{out}/R_{sid}$ is above one at the AGS, and it is 
relatively close to one at the SPS,\cite{heinz}-\cite{na49}, 
the data at RHIC gives an even 
smaller value.  This suggests that at RHIC this ratio reflects the opacity of
the emitting surface.
The origin of the smaller value of 
$R_{out}/R_{sid}$ for the opaque case is due to 
two effects.  First, the system radiates from a smaller geometrical 
region than in the transparent case.  Second, the emission itself
is preferentially from the region of the cylinder closest to the
radial vector made by ${\bf R}_{out}$, where this
vector is a radial vector in the direction of the pair transverse momentum
${\bf K}_T$.  This is because radiation from this
region is normal to the surface, and there is a factor of $\cos \phi$ 
associated with the flux in the direction of ${\bf R}_{out}$, where
$\phi$ is the angle between ${\bf R}_{out}$ and a normal
vector to the surface.  Finally there is also the emission 
from the decoupling volume which occurs at a well defined time, which 
we shall see does not significantly contribute to the flux of particles, 
although it still contributes to the {\it averaging} over different 
emission radii.

The case of $1+1$ dimensional longitudinal expansion is more complicated.
In later sections, we shall use a simple 
model of the matter, where there is a real first order
phase transition between a parton gas (gluons and massless quarks) 
and a pion gas.  The energy is emitted from the partonic  
phase.  This quark and gluon matter is assumed to be directly converted
into a flux of pions with the same energy and a blackbody distribution
at the temperature of emission.  This energy conservation condition
allows us to directly take a flux of gluons and quarks and convert it
into a spectrum of pions.  It of course will mess up details of
the fragmentation, and generates some increase in entropy, but for our purposes
such a crude treatment is sufficient to demonstrate the physics 
of opacity in a semi-quantitative way.  
The emission of the hadron degrees of freedom
is taken to be entirely pions.
We will find that the dominant emission at RHIC energies comes 
from times after the beginning of the mixed phase, but there is a significant 
mixture of gluonic and quark radiation which must convert into pions.

The characteristic time scale for emission
is of the order of the radius of the system before the radiation
is significantly attenuated.  This time scale is of the order of the 
size of the system for RHIC energies. Our model 
which incorporates this physics is described 
in detail in Sec. 4.  In the fifth section, we describe some of the details
of the computation of the HBT correlation function.

In the final section, we discuss the limitations of our approach.
The severest is that if we try to predict
the values of HBT radii as a function of $K_T $ of the pair, we do
not get the correct transverse momentum dependence.  We do however
get the correct $K_T$ dependence for $R_{out}/R_{sid}$.  


\section{The Covariant Current Ensemble Inputs}

To compute the emitted spectrum and the two particle distribution function, 
leading to the interferometry relations in which we are interested,
we consider the Covariant Current Ensemble formalism, \cite{ccef},\cite{pgg}.
In this formalism, the two particle correlation function can be written as
\beq
	C(k_1,k_2) = {{P_2(k_1,k_2) } \over {P_1(k_1)P_1(k_2)}} 
= 1 + {{\mid G(k_1,k_2)} \mid^2 \over {G(k_1,k_1) G(k_2,k_2)}}
\label{ck1k2}\eeq
where $P_1(k_i)$  and $P_2(k_1,k_2)$ are, respectively, the single particle
distribution and the probability for simultaneous observation of 
two particles with  momenta $k_1$ and $k_2$. The average momentum of the pair 
is defined as $ {\bf K} = ({\bf k_1} + {\bf k_2})/2$, and the relative 
momentum, as  ${\bf q} = {\bf k_1} - {\bf k_2}$. 

	The complex amplitude, $G(k_1,k_2)$, is written as 

\beq
G(k_1,k_2) =  \int d^4p \int d^4x \; e^{iq^\mu x_\mu}\;  D(x,p) \;  
j_0^\star(\frac{k_1.p}{m}) j_0(\frac{k_2.p}{m})
\; \; ,  \label{G12_1}\eeq 
and the single-inclusive distribution, $G(k_i,k_i)$, is written as

\beq
G(k_i,k_i) =  \int d^4p \; \tilde{D}(0,p) \;  |j_0(\frac{k_i.p}{m})|^2 
= \frac{d^3{\cal N}}{d^2k_i dy}= E_i \frac{d^3{\cal N}}{d^3k_i}  = P_1(k_i)
\; \; .  \label{Gii1}\eeq 

We see that 
\beq
\tilde{D}(q,p) =  \int d^4x \; e^{iq^\mu x_\mu}\;  D(x,p)
\; \; , \label{Dxp}\eeq 
where $D(x,p)$ is the normalized phase-space distribution at the 
instant of the emission.  

One should note that there is some arbitrariness in
how we decompose into $D(x,p)$ and the densities $j_0$.
We shall follow the original proposition in \cite{ccef},\cite{pgg}, and 
choose to include the thermal aspects of the collision in $j_0$, and 
the geometrical plus collective effects in $D(x,p)$. In any case 
both will be defined for the problems at hand.

There are several different possibilities for $D(x,p)$:

The first
is radiation from a transparent cylinder which is not expanding in the 
transverse directions and has no longitudinal expansion.
\beq
D(x,p) = g_p \; \kappa 
\;   \exp(-t^2/2\Delta t^2) \;    
 \delta(E_p - p^0) \; \delta^2({\bf p_T})\delta(p_z) \;  
\; \; . \label{Dxp2a}
\eeq

In this expression $p$ is to be interpreted as the momentum
of a particle at rest in the matter from which it is emitted,
and reflects the collective motion of the system. 
For matter at rest, ${\bf p} = 0$. The factor $g_p$ counts the number
of degrees of freedom of particle being emitted.  The factor
of $e^{-t^2/2\Delta t^2}$ is a Gaussian parameterization of the
source emission rate, so that the characteristic emission time
is $\Delta t$.    
(The blackbody rate emerges naturally in this picture, and 
$\kappa$ is a factor which modifies beyond the blackbody rate). 

The second case is emission from the surface of a cylinder without
longitudinal expansion.  Here
we must account for the fact that the particle is emitted with
a flux factor which depends upon the angle $\phi$ between the normal
vector to the surface and the direction of the pair transverse
momentum $\vec{K_T}$.  There is
also a factor of $\delta(r_T-R_T)$ which requires that particles are emitted
from the surface and a factor of $\Theta(\cos \phi)$ which requires they come
from the side of the surface from which they are emitted.
\beq
D(x,p) = g_p \; \kappa 
\;   \exp(-t^2/2\Delta t^2) \;  \delta(r_T-R_T) \; 
 \delta(E_p - p^0) \; \delta^2({\bf p_T}) \; \delta(p_z) \cos \phi \; 
 \Theta(\cos \phi) 
\; \; . \label{Dxp2b}
\eeq

Finally, there is the case of 
emission from the surface of a cylinder with longitudinal expansion  

\beq
D(x,p) = g_p \; \kappa 
 \;  \delta(r_T-R_T) \; 
 \delta(E_p - p^0) \; \delta^2({\bf p_T}) \; 
\delta(y-\eta) \cos \phi \; \Theta(\cos \phi) 
\; \; , \label{Dxp2c}
\eeq
 The factor of $\delta(y-\eta)$ is
the correlation between the velocity and coordinate of the emitting surface 
built into the Bjorken model \cite{Bj}.  The factor of 
$\kappa$ controls
the rate of emission from the surface as a function of time.  We will fix
this by requiring the system to be either a Quark Gluon Plasma, a hadron
gas or a mixture, with blackbody radiation and the mixture determined
by thermodynamics.

The currents correspond to thermal distributions and are written as 

\beq
\j_0(\frac{k_i.p}{m}) = \sqrt{u^\mu k_{i_\mu}} \;
\exp\{{-\frac{u^\mu k_{i_\mu}}{2 T(\tau)}}\} 
\; \; . \label{j0}
\eeq
Here $u^\mu = p^\mu/m$ is the four vector flow velocity
which is the four velocity of the emitting surface.  In the case where 
the currents are connected to the models illustrated by Eq. 
(\ref{Dxp2a})-(\ref{Dxp2b}) above, $u^\mu = (1, {\bf 0})$, 
i.e., 
associated to the absence of expansion along the longitudinal 
and transverse directions in those cases.   
We indicate by $T(\tau)$ the dependence of the temperature 
in the proper time, $\tau$, prior to the beginning of the mixed phase, 
as given by Eq.(\ref{temp}), in Section IV.

In the ideal Bjorken picture with no transverse flow, 
the 4-velocity $u^\mu$ and the momentum of the emitted particle, $k^\mu$, 
can be written as
\beq
u^\mu = (\cosh\eta, 0, 0, \sinh\eta) \; ; \; 
k^\mu = (m_T \cosh y, {\bf k_T}, m_T \sinh y)
\; \; . \label{umu}\eeq

\bigskip

\section{A Very Simple Model for 
R$_{\bf \lowercase{out}}/$R$_{\bf \lowercase{sid}}$}

We first work out the simplest model for emission so that we can get some 
conceptual understanding of the physics involved in the various cases.  
We shall consider a system without radial flow. The approximations will 
be fixed in the next section when we consider a more realistic 
model which we compare with data.

First consider emission from a transparent cylinder.
A little algebra gives
\beq
	C_{\tiny out} = 1 + \exp\left[-\frac{K_T^2 q_{\tiny out}^2 \Delta t^2}
	{(K^2+m^2)}\right] 
\left[  2J_1 (q_{\tiny out} R_T) / (q_{\tiny out} R_T) \right] ^2
\ee
and 
\beq
	C_{\tiny sid} = 1 + \left[  2J_1 (q_{\tiny sid} R_T) / 
	(q_{\tiny sid} R_T) \right] ^2
, \eeq
where $J_1(x)$ is the Bessel function of the first kind and order one.

In this equation, it is absolutely clear that $R_{out} > R_{sid}$,
and that if $\Delta t \sim R$, then $R_{out} \sim 2 R_{sid}$

Now let us consider the result for surface emission.  In this case,
a little algebra yields
\beq
	C_{\tiny out} =  1 + \exp\left[-\frac{K_T^2 q_{\tiny out}^2 \Delta t^2}
	{(K^2+m^2)}\right] \mid I(q_{\tiny out}R_T) \mid^2
\eeq
where
\beq
	I(x) = {1 \over 2} \int_0^\pi d\phi \sin(\phi) \exp[i x \sin(\phi)]
\eeq
On the other hand,
\beq
       C_{\tiny sid} =  1 +\mid {{\sin(q_{\tiny sid} R_T)} 
       \over (q_{\tiny sid} R_T)} \mid^2
\ee
In Fig. 1, we plot $C_{sid}$ and, for stressing the geometrical differences 
only, we ignored the time contribution to $C_{out}$, by fixing $\Delta t =0$ 
in the plots. Fitting the curves by Gaussian distributions, we find that, in the 
opaque case, $R^{eff}_{sid} \approx R_T \times 0.53 $ 
and $R^{eff}_{out} =R_T \times 0.22$. In the transparent case, the fit 
results in an effective radius of $R^{eff}_{out} \approx 0.61 R_T$.

The basic result we get from this analysis is that for the opaque cylinder,
unlike the transparent one, we can easily have $R_{out} < R_{sid}$.  In fact
for $R_{out} \sim \Delta t (K_T/E_K)$ (since the $R_T$ contribution is very 
small), the ratio of $R_{out} /R_{sid} \sim \Delta t (K_T/E_K)/ R_T$ 
unlike the transparent cylinder case where $R_{out}/R_{sid} \sim 
\sqrt{\Delta t^2 (K_T/E_K)^2 + R_T^2 }/ R_T$.

\bigskip

\section{A Simple Model for R$_{\bf \lowercase{out}}/$R$_{\bf \lowercase{sid}}$}

In this model, we incorporate longitudinal expansion.  Here the system
cools as it expands, so we need to have the rate of emission vary as a function
of time.  

We need to have dynamical description of the microphysics to be able
to do this. Our more or less conventional model is a  
Quark-Gluon Plasma (QGP) phase at some temperature above the critical 
temperature, $T_0 > T_c$.  
We choose this critical temperature $T_c = 175$ MeV 
to be consistent with lattice Monte-Carlo data. 
The system goes into a mixed phase of QGP and hadron gas at the
critical temperature, and we take this hadronic gas to be 
composed of an ideal gas of pions.  Below $T_c$, it is only an ideal 
gas of pions. While in the low temperature pionic phase, 
the system cools until it reaches a decoupling temperature, $T_f$. 
We choose $T_f = 150~$MeV 
to be consistent with the typical energy per particle
observed in the RHIC experiments, and to fit the observed $p_T$ distributions
of pions. However, these values chosen for $T_c$ and $T_f$ are not crucial for 
qualitatively  reproducing the results we discuss here, since these
results are only weakly sensitive to these particular values.

The system produced in a heavy ion collision will expand and the temperature 
will gradually decrease. The initial expansion is in the Quark-Gluon Plasma 
phase, and the system expands longitudinally.
After this initial stage, lasting  
($\tau_c - \tau_0$), the transition temperature, $T_c$, is reached  
and the evolution continues in the mixed phase, during which the 
temperature remains constant with time. 
The mixed phase continues for a longer 
period, ending after an elapsed interval ($\tau_h - \tau_c$). 
Then, the system converted into a gas of pions, further expands until 
the decoupling temperature, $T_f$, is reached. 
At this point, the system is quite dilute, and much 
of the particles have been evaporated from its surface. Thus, we 
consider that, once  $T_f$ is reached, the system decouples in 
an instantaneous volumetric emission.

The Bjorken hydrodynamical model\cite{Bj}  should be able to describe 
the system during its  evolution from  formation until the time it 
breaks up.  We will supplement this
with radiation from the surface of the matter.  We will take the 
radius at which this radiation takes place to be a constant and 
equal to the radius of the nuclei.  
We only consider impact parameter zero collisions.  
In the last stage of the system evolution we consider a volumetric 
emission at freeze-out but, as in the surface emission, no transverse 
flow was introduced in the computation. 
This could be one of the main reasons 
for obtaining a much weaker $K_T$ decrease of the transverse radii, as 
compared to RHIC data\cite{star,phenix}.

Another ingredient in our model is the hypothesis that the system 
will emit from its external surface similarly to a  blackbody, 
starting shortly after being formed, at $\tau = \tau_0$. In this way, 
quark and gluon degrees of freedom have to be considered 
in the QGP and mixed phases. The hadronic degrees of freedom should, 
in principle, include a complete set of resonances later decaying into 
pions. However, in this initial description, and for the sake of simplicity, 
we will consider a hadronic gas constituted of only pions.  No complex 
mechanism for the QGP hadronization will be considered 
in detail at this point, although hadronization must take place. 
In other word, in first approximation, we will consider the evaporation 
of ``gluons'' and ``quarks'' (as hadronized pions) from the external 
surface of the system in the same way as emission of pions, except for 
the number of degrees of freedom. 

We can estimate the emitted energy as well as the total entropy 
associated to each stage. 
In the initial phase, lasting from $\tau_0$ to $\tau_c$, 
we can estimate the emitted energy as a 
function of time considering the emission by an expanding cylinder 
of transverse radius $R_T$ and length $h$, in a certain time interval 
between $\tau$ and $\tau+d\tau$ by 
\beq
dE_{in} = - \kappa \sigma T^4 \; 2 \pi \; R_T \; h \; d\tau - 
\frac{1}{3} \; \sigma T^4\;  \pi \; R_T^2  dh
\; \; , \label{denergy}\eeq
where the first term comes from the blackbody type of energy  
radiated from the surface of the cylinder, and the second term 
results from the mechanical work due to its expansion. The $\kappa$
factor was introduced to take into account that the system has some opacity 
to surface emission.  The constant $\sigma$ is 
is proportional to the number of degrees  of freedom in the system. 

	By integrating Eq. (\ref{denergy}) we get for the energy density 
(i.e., $\epsilon = E/V$) 
\beq
\epsilon_{in} = \epsilon_0 (\frac{\tau_0}{\tau})^\frac{4}{3} 
e^{-\frac{2 \kappa}{R_T}(\tau-\tau_0)}
\; \; . \label{energy_in}\eeq

From the above expression we see that we obtain an extra multiplicative 
factor, $e^{-\frac{2 \kappa}{R_T}(\tau-\tau_0)}$, in addition to that coming 
from the Bjorken picture. Remembering that, in the Bjorken picture, the 
relation between the energy density, $\epsilon$, and the proper-time, $\tau$, 
is given by $\epsilon/\epsilon_0 = (\tau_0/\tau)^{4/3}$, 
and that $\epsilon = \sigma T^4$ 
for a blackbody-type radiation, then the variation of 
the temperature in the initial stage, i.e., prior to the beginning 
of the phase transition,
 follows immediately as
\beq
T(\tau) = T_0 (\frac{\tau_0}{\tau})^\frac{1}{3} 
e^{-\frac{\kappa}{2 R_T}(\tau-\tau_0)}
\; \; . \label{temp}\eeq

	In order to fix the initial conditions of the 
evaporating and expanding fireball, we follow the evolution of 
the entropy, using the observed final particle multiplicity 
density $dN/dy$ and its relation to the final entropy density $dS/dy$ 
as a constraint.  We decompose the total entropy $S_{tot}$ into 
its contributions $S_{in}$ from the fireball interior and 
from the emitted particles $S_{emit}$, all of which 
are functions of the proper-time.  During the first stage, 
from $\tau_0$ to the beginning of the mixed phase at time 
$\tau_c$ we, can write 
\beq
s_{in} = \frac{S_{in}}{V} = \frac{1}{T}(\epsilon+p)_{in} = 
s_0 (\frac{\tau_0}{\tau})e^{-\frac{3 \kappa}{2R_T}(\tau-\tau_0)}
\rightarrow S_{in} = S_0 e^{-\frac{3 \kappa}{2R_T}(\tau-\tau_0)}
\; \; , \label{entropy_in}\eeq
where $s_0 = \frac{4}{3} (\sigma\epsilon_0^3)^{1/4}$ is the initial 
entropy density. In the second step of Eq. (\ref{entropy_in}) 
we used that, for the Bjorken type of longitudinal expansion, the volume 
$V(\tau)$ increases linearly with time. 

	On the other hand, the entropy associated with the emission can 
also be estimated. For this, we write the element of emitted 
entropy, $d S_{emit}$, in the interval between $\tau$ and 
$\tau + d \tau$ as proportional to the 
entropy on the surface (i.e., to the entropy density times the surface 
area, $2 \pi R_T \tau$). The emissivity, $\kappa$, is the proportionality 
constant. If we divide both sides by the corresponding volume, 
$V = \pi R^2_T \tau$, we then get  

\beq
\frac{dS_{emit}}{V} = \frac{2 \kappa}{R_T} s_{in} d\tau 
= \frac{2 \kappa}{R_T} s_0 (\frac{\tau_0}{\tau}) 
e^{-\frac{3 \kappa}{2R_T}(\tau-\tau_0)} d\tau
\rightarrow S_{emit} = \frac{4}{3} S_0 [1 - 
e^{-\frac{3 \kappa}{2R_T}(\tau-\tau_0)}]
\; \; . \label{entropy_emit1}\eeq

This results in the total entropy of the initial stage as being 

\beq
S_{tot} = \frac{4}{3} S_0 - \frac{1}{3} S_0 \;
e^{-\frac{3 \kappa}{2R_T}(\tau-\tau_0)}
\; \; , \label{entropy_tot1}\eeq
so that, at $\tau=\tau_0 \rightarrow S_{tot} = S_0$.  
Note that this result
requires that there be entropy produced during the emission from the
surface with the added input that the temperature changes
as a function of time. In the mixed phase there 
will be no such effect, since there the emitted quanta are 
simply the number of the quanta in the system, and no entropy
increase in particle number is generated.  Note that the entropy
for a massless gas is directly proportional to particle number.  
The situation is simply different 
when one has an expanding system with a variable temperature.                        
                                                                      
	During the phase transition, the energy and 
entropy can be estimated similarly, leading to 

\beq
\tilde{E}_{emit} = \tilde{E_c} (1 - 
e^{-\frac{2 \kappa}{R_T}(\tau-\tau_c)}) \; ; \; 
\tilde{S}_{in} = \tilde{S}_c \;
e^{-\frac{2 \kappa}{R_T}(\tau-\tau_c)} \; ; \;
\tilde{S}_{emit} = \tilde{S}_c (1 - 
e^{-\frac{2 \kappa}{R_T}(\tau-\tau_c)})
\; \; . \label{ener_phentr_ph}\eeq
These results follow simply from Eq. (\ref{denergy}) when the second term
on the right hand side is set to zero.  During the mixed phase no work
is done in expansion because the temperature does not change and
the expansion conserves entropy.  Therefore only the first term on the 
right hand side of Eq. (\ref{denergy}) contributes, and it is easy to 
integrate, since it is $2 \kappa E d\tau /R$. 

From the above relations we can see that, during the phase transition, 
the total entropy, $\tilde{S}_{tot} = \tilde{S}_{in}+\tilde{S}_{emit} = 
\tilde{S}_{c}$ is conserved. During this extended period, the temperature 
remains constant with time ($T = T_c$), so that the previous 
relation in Eq. (\ref{temp}) no longer holds.  
Note that there is a difference in the exponential behavior in the mixed phase
relative to that in the QGP phase.  This is a consequence of the fact that 
the temperature is time dependent in the QGP phase, but time independent
in the mixed phase, so that in the first
case the rate of emission from the surface is different.

During the 
phase transition the system is in a mixed phase of Quark-Gluon Plasma
and hadronic gas. If the fraction of the fluid in the QGP phase is $f$, 
then in the hadronic (pion) phase it would be ($1-f$). On the other hand, 
from Eq. (\ref{ener_phentr_ph}), we see that the portion of the entropy 
density still in the system is given by $\tilde{s}_{in}= \tilde{s}_c 
(\frac{\tau_c}{\tau}) \exp{[-\frac{2\kappa}{R_T}(\tau-\tau_c)]}$. Consequently, 
$\tilde{s}_{in} = f \; \tilde{s}_{QGP} + (1 - f) \; \tilde{s_h}$, where 
$\tilde{s}_{in}(\tau_c)=\tilde{s}_c$ and $\tilde{s}(\tau_h)=\tilde{s}_c 
(\frac{\tau_c}{\tau_h}) \exp{[-\frac{2\kappa}{R_T}(\tau_h-\tau_c)]}$
. If we substitute these expressions into that for $\tilde{s}_{in}$, 
we then get 

\beq
f = \left( \frac{\tau_h \; e^{-\frac{2 \kappa}{R_T}(\tau-\tau_c)}- 
\tau \; e^{-\frac{2 \kappa}{R_T}(\tau_h-\tau_c)}}
{\tau_h - \tau_c \; e^{-\frac{2 \kappa}{R_T}(\tau_h-\tau_c)}} \right) 
\frac{\tau_c}{\tau} \;  \;  \;\;  \;  ;  \;  \; \;  \;  \;
(1 - f) = \left( \frac{\tau- \tau_c \; e^{-\frac{2 \kappa}{R_T}(\tau-\tau_c)}}
{\tau_h - \tau_c \; e^{-\frac{2 \kappa}{R_T}(\tau_h-\tau_c)}} \right) 
\frac{\tau_h}{\tau} 
\; \; . \label{fractions}\eeq

Finally, we need to estimate the initial 
values $T_0$ and $\tau_0$, as well as the proper time, $\tau_c$, 
corresponding to the on-set of the phase transition. We estimate $\tau_0$  
by means of the Uncertainty Principle, i.e., 
$\langle E_0 \rangle \tau_0 \approx \hbar$, and 
by  

\beq
\langle E_0 \rangle = \frac{\int dp \; p^3 \; e^{-p/T_0}}
{\int dp \; p^2 \; e^{-p/T_0}} = 3 T_0
\; \; , \label{E0}\eeq
from which we easily determine the initial time as

\beq
\tau_0 \sim \frac{\hbar}{3 T_0} = \frac{0.197}{3 (T_0/{\rm GeV})} {\rm fm}/c 
\; \; . \label{tau0}\eeq

On the other hand, the constraint on $T_0$ and $\tau_0$ has to come 
from the experiment. At RHIC, the average produced pion multiplicity 
per unit of rapidity is ${\cal N} \sim 1000$, 
which should be proportional to the initial entropy, 
$S_0$, i.e., 

\beq
S_0 = \Gamma {\cal N} = \left[ (g_{\small g} + g_{\small q}) \times
(\frac{4}{3}) \frac{\pi^2}{30} T_0^3 \right] \pi R_T^2 \; \tau_0
\; \; . \label{N}\eeq

In the above expression, we have related the initial entropy, $S_0 = s_0 V_0$ 
to the initial entropy density and the initial volume of the system, 
$V_0 = \pi R_T^2 \tau_0$, this last one estimated in the Bjorken fashion. 
The degeneracy factors, $g$, are given by the gluon 
degrees of freedom, $g_{\small g} = 2 (spin) \times 8 (color) $, and 
the quark/anti-quark degrees 
of freedom, $g_{\small q} = \frac{7}{8}[2 (spin) \times 2 (q+\bar{q}) 
\times 3  (color) \times N_f (flavor)]$, 
which add up to $g_{\tiny qgp} = g_{\small g} + g_{\small q}$. 
In the case of pions, the degeneracy factor is $g_{\small \pi} = 3$.

From Eq.(\ref{tau0}) and (\ref{N}) we can determine $T_0$ as 
\beq
T_0 = \sqrt{{\cal N} \Gamma}
\left[ (g_{\small g} + g_{\small q}) \times 
\frac{4 \pi^3}{270} \frac{R_T^2}{(0.197)^2} 
\right]^{-1/2} {\rm GeV}
\; \; . \label{temp0}\eeq
As an example, if we take $\Gamma = 3.6$, as estimated by the entropy per 
particle ($S_\pi/N_\pi$) of a pion gas at freeze-out, 
then $T_0 \sim 411$ MeV and $\tau_0 \sim  0.160$ fm. 

For estimating the instant corresponding to the beginning of the mixed 
phase, $\tau_c$, we consider Eq.(\ref{temp}) at $\tau = \tau_c$, 
resulting in 

\beq
\tau_c \; e^{\frac{3 \kappa}{2 R_T} \tau_c} = 
\left( \frac{T_0}{T_c} \right)^3 \; \tau_0 \; 
e^{\frac{3 \kappa}{2 R_T} \tau_0} 
\; \; , \label{tauc}\eeq
which can be numerically estimated for fixed values of $\kappa$ 
and $T_c$ (for this, we will consider $T_c = 175$ MeV). 

In order to estimate the instant corresponding to the end of the mixed 
phase, $\tau_h$, we need to know $\tilde{s}_c$ and $\tilde{s}_h$. For the 
first one, we consider a system of gluons and (massless) quarks forming 
an ideal gas, resulting in $\tilde{s}_c=(g_g + g_q)\frac{2 \pi^2}{45} T_c^3$. 
For estimating $\tau_h$, we consider the pions as massless particles, while 
in the system, leading to $\tilde{s}_h=g_\pi \frac{2 \pi^2}{45} T_c^3$. 
Then, equating the expression for $\tilde{s}_{in}$ at $\tau=\tau_h$, we obtain

\beq
\tau_h \; e^{\frac{2 \kappa}{R_T} \tau_h} = 
\left( \frac{g_g+g_q}{g_\pi} \right) \tau_c \; 
e^{\frac{2 \kappa}{R_T} \tau_c} 
\; \; , \label{tauh}\eeq
which can be estimated numerically for a fixed value of $\kappa$, since 
$\tau_c$ was already determined by Eq. (\ref{tauc}).

We assume that, at the end of the phase transition, corresponding to 
$\tau=\tau_h$, the system is an ideal gas of pions 
(no resonances are considered in this initial estimate), 
which continues to expand and cool down, until the temperature drops to 
$T_c = 150$ MeV, 
corresponding to an instant $\tau=\tau_f$. At this point, whatever is remnant 
of the system breaks up. The portion of the pions still in the system at 
that time we call {\large ${\cal V}$}, i.e., the fraction of the system that is 
emitted from the entire volume at the time $\tau_f$. 
This decoupling instant can be estimated by an 
expression similar to Eq.(\ref{temp})), with $T_0$ replaced by $T_c$, and 
$\tau_0$ by $\tau_c$, resulting in

\beq
\tau_f \; e^{\frac{3 \kappa}{2 R_T} \tau_f} = 
\left( \frac{T_c}{T_f} \right)^3 \; \tau_h \; 
e^{\frac{3 \kappa}{2 R_T} \tau_h} 
\; \; , \label{tauf}\eeq
which can be numerically estimated for fixed values of $\kappa$, $T_c$, and 
$T_f$, and corresponding $\tau_h$.

We illustrate in Table 1 below these variables for two different 
assumptions on the emissivity, $\kappa$. The values of the temperatures 
considered were $T_0 \sim 411$ MeV (obtained from 
Eq. (\ref{temp0})), $T_c = 175$ MeV and $T_f=150$ MeV. We also include 
the corresponding estimates of the fraction of particles emitted 
from the surface, ${\cal S}$,  in the interval $\tau_0 \le \tau \le \tau_f$, 
as well as the fraction from the volumetric emission, {\large ${\cal V}$}, 
at $\tau=\tau_f$.

To estimate the fraction of the particles emitted from the surface, 
${\cal S}$, and from the volume, {\large ${\cal V}$}, we proceed as 
follows, estimating the contribution from each stage of the system evolution. 
For that, we estimate the emitted entropy in each stage, similarly to the 
procedure described in Eq.(\ref{entropy_in})-(\ref{ener_phentr_ph}). 

The fraction of the input particles, ${\cal N}$, emitted in 
the first period, $\tau_0 \le \tau \le \tau_c$, is given by  

\beq
\frac{{\cal N}_1}{{\cal N}} = \frac{4}{3}\; (1-e^{-\frac{3 \kappa }{2 R_T}
(\tau_c-\tau_0)})
\; \; . \label{Ntau0tauc}\eeq

	Similarly, the fraction  ${\cal N}_2/{\cal N}$ emitted during the 
phase transition, $\tau_c \le \tau \le \tau_h$, can be written as

\beq
\frac{{\cal N}_2}{{\cal N}} = e^{-\frac{3 \kappa }{2 R_T}
(\tau_c-\tau_0)} (1-e^{-\frac{2 \kappa}{R_T}(\tau_h-\tau_c)})
\; \; . \label{Ntauctauh}\eeq

	The fraction emitted during the pure pion phase, 
$\tau_h \le \tau \le \tau_f$, up to reaching $T_f$, can be estimated by 
 
\beq
\frac{{\cal N}_3}{{\cal N}} = \frac{4}{3}\; e^{-\frac{3 \kappa }{2 R_T}
(\tau_c-\tau_0)}\; e^{-\frac{2 \kappa}{R_T}(\tau_h-\tau_c)}
(1-e^{-\frac{3 \kappa }{2 R_T}(\tau_f-\tau_h)}) 
\; \; . \label{Ntauhtauf}\eeq

	Then, the fraction of the particles emitted from the surface up to 
$\tau = \tau_f$ is 
	 
\beq
\frac{{\cal S}}{\cal N} = \frac{1}{\cal N}({\cal N}_1 + {\cal N}_2+ {\cal N}_3) 
\; \; . \label{Surf}\eeq

	Finally, the remnant fraction at $\tau= \tau_f$, emitted from the 
entire volume, is given by
	 
\beq
\frac{{\large {\cal V}}}{\cal N} = \frac{{\cal N}_4}{{\cal N}} = 
e^{-\frac{3 \kappa }{2 R_T}
(\tau_c-\tau_0)}\; e^{-\frac{2 \kappa}{R_T}(\tau_h-\tau_c)}
e^{-\frac{3 \kappa }{2 R_T}(\tau_f-\tau_h)}
\; \; . \label{Ntauf}\eeq

As a result, the ratio of the total number of emitted particles, 
${\cal N}_{\bf tot}$, with respect to the input number, ${\cal N}$, 
is given by

\beq
\frac{{\cal N}_{\bf tot}}{{\cal N}} = \frac{1}{{\cal N}}
({\cal N}_1 + {\cal N}_2+ {\cal N}_3 + {\cal N}_4) 
\; \; . \label{Ntot}\eeq

\vskip 1.cm
\bigskip
\begin{center}
{\bf TABLE 1: Values of the proper-time parameters $\tau_0$, $\tau_c$, 
$\tau_h$, $\tau_f$, as well as of the surface, ${\cal S}$ 
(for $\tau_0 \le \tau \le \tau_f$), and 
the volume,  {\large ${\cal V}$} (at $\tau_f$), 
emitted fluxes, for two values of the emissivity, $\kappa$.} \\
\end{center}

\vskip 1.cm

{\large

\begin{center}
\begin{tabular}{|c|c|c|c|c|c|c|}
\hline
  $\kappa$  & $\tau_0$  & $\tau_c$  &  $\tau_h$  &  $\tau_f$ &  
  ${\cal S}/{\cal N}$  & 
 {\large ${\cal V}$}$/{\cal N}$ \\ 
 &  (fm/c)  &  (fm/c) &   (fm/c)    & (fm/c)  & 
 ($\tau_0 \le \tau \le \tau_f$) & (at $\tau_f$) \\ \hline\hline
 ~ 1 ~ & ~ 0.160  ~ & ~ 1.54 ~ & ~ 5.73 ~ & ~ 6.97 ~ & 0.844 & 0.156 \\ \hline
 ~ 0.5 ~ & ~ 0.160 & ~ 1.75 ~ & ~  8.37 ~ & ~ 10.5 ~ & 0.758 & 0.242 \\ \hline
\end{tabular}
\end{center}
}
\bigskip
\bigskip


As we can see from the fractions in Table 1 above, in our model there is a 
small increment ($\sim 7.6-10 \%$) in the total number of particles 
with respect to the initial one. 
We show in Fig. 2 the evolution of the emitted flux with proper time, 
by plotting the fractions normalized to the total number of produced pions, 
as a function of $\tau$. 
The two different cases correspond to different surface emissivities. The 
curves end at the decoupling temperature.  In the case of the greatest rate
of surface emission, about $84\%$ of the total radiation comes from the surface
and the rest from the decoupling volume.  It is about $76\%$ for the other 
case we show for illustration.  In the first case, the system decouples at 
a time of about $7$ fm/c and in the second case at about $10.5$ fm/c.  
We assume a decoupling temperature of $150$ MeV. 
We should note that the decoupling proper-times within our model 
are significantly shorter than those in hydrodynamics. 
This is mainly due to the fact that we allowed a 
higher surface emissivity than those kind of models, which may be 
the key to explain the proper-times observed at RHIC, which are  smaller 
than expected from hydro predictions.

\bigskip\bigskip\bigskip

\section{single- and two-particle probability distributions}
\bigskip

	We define the average momentum of the pair as 
$K=\frac{1}{2}(k_1+k_2)$ and the relative momentum as $q=(k_1-k_2)$. 
They satisfy $q^\mu K_\mu = 0$, and, consequently, 
the temporal component of $q^\mu$ can be written as 
$q^0 = \frac{\bf q.K}{K^0}$. In the limit that is interesting for 
interferometry, we can consider $|{\bf q}| \ll |{\bf K}|$, which implies 
that $K^0 \approx \sqrt{|{\bf K}|^2 + m^2} = E_K$. 

	In both expressions for $C(k_1,k_2)$  and $C(k_i,k_i)$, we see that, 
due to the form of the phase-space distribution in Eq.(\ref{Dxp2c}), 
integration over the variables involving delta functions are straightforward. 
We should remember that, due to the factor $\Theta(\cos \phi)$ in 
Eq.(\ref{Dxp2c}), the integration over $d\phi$ runs in the interval 
$[-\pi/2,\pi/2]$, while the limits on the proper time integration 
would be $[\tau_0,\tau_f]$. However, as the system has different composition 
in each phase, we should split this time integration to be $[\tau_0,\tau_c]$,  
$[\tau_c,\tau_h]$, and then $[\tau_h,\tau_f]$.
The rapidity integration should run, in the 
Bjorken picture, from ($-{\small \infty},+{\small \infty}$). 

	We recall that $f$ and $(1-f)$ are, respectively, the fractions of 
the system in the QGP phase and in the hadronic phase, according to the 
expression given in Eq.(\ref{fractions}). Then, by 
taking into account the above observations, the expression for the 
complex amplitude, after some algebraic manipulation, can finally be 
written as 


\bea
\!\!G(k_1,k_2) \; & &  \propto 
\kappa R_T
\int_{-\frac{\pi}{2}}^{\frac{\pi}{2}} d\phi \cos\phi 
\int_{-{\small \infty}}^{+{\small \infty}} dy [E_K \cosh y - K_L \sinh y] 
\nonumber
\\
& &
e^{- i q_T R_T \cos{(\alpha - \phi)} }
\{  g_{\tiny qgp} \int_{\tau_0}^{\tau_c} \tau d\tau 
e^{ i \tau [\frac{K_T}{E_K} q_T \cos\alpha \cosh y + 
q_L (\frac{K_L}{E_K} \cosh y - \sinh y )]} 
\nonumber
\\ & & 
\exp \left[ - \frac{1}{T_0}(E_K \cosh y - K_L \sinh y) 
(\frac{\tau}{\tau_0})^{1/3} 
e^{\kappa(\tau-\tau_0)/(2 R_T)} \right] + 
\nonumber
\\ & & 
g_{\tiny qgp} \; \int_{\tau_c}^{\tau_h} \tau \; d\tau \;
( f ) \;
e^{ i \tau [\frac{K_T}{E_K} q_T \cos\alpha \cosh y + 
q_L (\frac{K_L}{E_K} \cosh y - \sinh y) ]}
e^{- \frac{1}{T_c}(E_K \cosh y - K_L \sinh y)} + 
\nonumber
\\ & & 
g_{\small \pi}\int_{\tau_c}^{\tau_h} \tau \; d\tau  \; (1 - f) \; 
e^{ i \tau [\frac{K_T}{E_K} q_T \cos\alpha \cosh y + 
q_L (\frac{K_L}{E_K} \cosh y - \sinh y) ]}
e^{- \frac{1}{T_c}(E_K \cosh y - K_L \sinh y)} + 
\nonumber
\\ & & 
g_{\small \pi}  \int_{\tau_h}^{\tau_f} \tau \; d\tau \; 
e^{ i \tau [\frac{K_T}{E_K} q_T \cos\alpha \cosh y + 
q_L (\frac{K_L}{E_K} \cosh y - \sinh y )]} 
\nonumber
\\ & & 
\exp \left[ - \frac{1}{T_c}(E_K \cosh y - K_L \sinh y) 
(\frac{\tau}{\tau_h})^{1/3} 
e^{\kappa(\tau-\tau_h)/(2 R_T)} \right] \} + 
\nonumber
\\& & 
\frac{g_{\small \pi}\tau_f}{\pi} 
\int_{0}^{2 \pi} d\phi \; \int_{0}^{R_T} \; r_T \; dr_T
\int_{-{\small \infty}}^{+{\small \infty}} dy [E_K \cosh y - K_L \sinh y] 
\nonumber
\\& & 
e^{ i \tau [\frac{K_T}{E_K} q_T \cos\alpha \cosh y + 
q_L (\frac{K_L}{E_K} \cosh y - \sinh y )]
- i q_T R_T \cos{(\alpha - \phi)}} 
e^{- \frac{1}{T_f}(E_K \cosh y - K_L \sinh y)} 
. \label{G12_2}\eea

The five terms composing the correlation function represent, respectively, 
the emission from the quark and gluon initial stage, their contribution 
during the mixed phase, the pion emission also from the surface in that phase, 
the emission during the pure pionic phase up to reaching the freeze-out 
temperature, $T_f$, and finally, the volumetric instantaneous decoupling 
once it was reached. 

\medskip
Similarly, we can write the spectrum, as
\medskip

\bea
\!\!G(k_i,k_i)\; & &  \propto  
\kappa R_T
\int_{-\frac{\pi}{2}}^{\frac{\pi}{2}} d\phi \cos\phi 
\int_{-{\small \infty}}^{+{\small \infty}}  dy  \;  
[E_i \cosh y - k_{i_L} \sinh y]
\nonumber
\\
& &
\{ \; g_{\tiny qgp} \int_{\tau_0}^{\tau_c} \tau  \; d\tau 
\exp [ - \frac{1}{T_0}(E_i \cosh y - k_{i_L} \sinh y) (\frac{\tau}{\tau_0})^{1/3} 
e^{\kappa(\tau-\tau_0)/(2 R_T)} ] + 
\nonumber
\\ & & 
g_{\tiny qgp}  \nonumber
\int_{\tau_c}^{\tau_h} \tau  d\tau \; 
( f ) \; 
e^{- \frac{1}{T_c}(E_i \cosh y - k_{i_L} \sinh y)}  +
g_{\small \pi}\; \int_{\tau_c}^{\tau_h} \tau  \; d\tau \;
(1 - f) \;
e^{- \frac{1}{T_c}(E_i \cosh y - k_{i_L} \sinh y)} + 
\nonumber
\\ & & 
g_{\small \pi} \int_{\tau_h}^{\tau_f} \tau  d\tau 
\exp [ - \frac{1}{T_c}(E_i \cosh y - k_{i_L} \sinh y) 
(\frac{\tau}{\tau_h})^{1/3} 
e^{\kappa(\tau-\tau_h)/(2 R_T)} ] \; \} + 
\nonumber
\\ & & 
\frac{g_{\small \pi}\tau_f}{\pi} 
\; \int_{0}^{2 \pi} d\phi \; \int_{0}^{R_T} \; r_T \; dr_T
\int_{-{\small \infty}}^{+{\small \infty}}  \; dy  \;  
[E_i \cosh y - k_{i_L} \sinh y]
e^{- \frac{1}{T_f}(E_i \cosh y - k_{i_L} \sinh y)} 
\; . \label{Gii_2}\eea

\bigskip

We remind that $G(k_i,k_i)$ is the spectrum as written in Eq.(\ref{Gii1}). 
If we then integrate separately the terms of Eq. (\ref{Gii_2}) in 
$d^3k_i/E_i$, i.e., in the 
intervals $\tau_0 \le \tau \le \tau_c$, $\tau_c \le \tau \le \tau_h$, 
$\tau_h \le \tau \le \tau_f$, and at $\tau = \tau_f$, we recover the 
number of emitted particles in each interval, as seen in 
Eq.(\ref{Ntau0tauc}), (\ref{Ntauctauh}), (\ref{Ntauhtauf}), and 
(\ref{Ntauf}), respectively, except for 
an overall normalization constant, which is cancelled when we estimate 
the interferometric relations, as in Eq. (\ref{ck1k2}). 

	For estimating the (real) amplitudes $G(k_i,k_i)$ in the  
denominator of Eq. (\ref{ck1k2}), we wrote ${\bf k_1} = {\bf K} + {\bf q}/2$  
and ${\bf k_2} = {\bf K} - {\bf q}/2$, 
from the definition of the momenta ${\bf K}$ and the relative 
momentum ${\bf q}$, since we are not generating 
the individual momenta and later averaging over all of them, as done 
in the experiment. In this first approach, that is the way we connected 
the momenta $k_i$ in the spectra $C(k_i,k_i)$ to the momenta appearing 
in the complex amplitude, $C(k_1,k_2) = C(q,K)$. 

This means that we can write 

\medskip
\beq
|{\bf k_T}_{1,2}| = \sqrt{({\bf K_T} \pm {\bf q_T}/2)^2} = 
\sqrt{{\bf K}^2_{\bf T} + \frac{{\bf q}^2_{\bf T}}{4} \pm K_T q_T \cos\alpha}
\; \; . \label{kt12}\eeq

\medskip
\beq
| k_{L_{1,2}}| =  \sqrt{K_L^2 + \frac{q_L^2}{4} \pm K_L q_L} \; \; ; \; 
E_{1,2} = \sqrt{m^2 + ({\bf K_T} \pm {\bf q_T}/2)^2 + (K_L \pm q_L/2)^2}
\; \; . \label{kl12}\eeq

\bigskip
In these equations, $\alpha$ is the angle between ${\bf K}_T$ and
${\bf q}_T$.

	Due to the azimuthal symmetry of the problem, we can choose 
${\bf K}_T$ along the $x$-axis, without any loss of generality. In this way, 
we see that the component of the relative momentum in the 
{\sl outward} direction (${\bf q_T} \parallel {\bf K_T}$ ), ${\bf q_{T_{out}}}$, 
will be along the $x$-axis, while the 
{\sl sideward} component (${\bf q_T} \perp {\bf K_T}$ ), ${\bf q_{T_{sid}}}$, 
will be directed along the $y$-axis, i.e., 
$q_{\small x} = q_{T_{out}}$ and $q_{\small y} = q_{T_{sid}}$. 
This implies that, in the first case, we chose $\alpha = 0$ and, 
in the second, $\alpha = \pi/2$. 

	In order to check how the spectra estimated within our model and 
the above discussed relations behave compared to data (PHENIX minimum bias
\cite{phenix2}), we 
plot the single-inclusive distribution in Fig. 3. From here on we limit 
our estimates and discussions  to the central rapidity region, i.e., 
$y_i = 0$ (which implies that $k_{i_L} = 0$, and, consequently, $K_L = 0$ 
and $q_L = 0$). In Fig. 3, we show the spectra, within a 
constant arbitrary normalization for the set of interrelated values shown in 
the first line of Table 1, and for two values of the emissivity, 
$\kappa=0.5, 1$. We see that both curves describe well the spectrum
in the low momentum region of the pions, up to roughly 
$k_{i_T}\approx 1$ GeV/c. We shall consider these possibilities also when 
studying the correlation function and fitted radii. 

	In Fig. 4 we show the correlation functions $C(q_{T_{out}})$ vs. 
	$q_{T_{out}}$ 
(solid curves), corresponding to $\alpha=0$, and $C(q_{T_{sid}})$ vs. 
$q_{T_{sid}}$ 
(dashed ones, very close to one another), for $\alpha=\pi/2$, in the 
same plot for better visualize the differences. When calculating the 
correlation function in terms of $q_{T_{out}}$, we fixed $q_{T_{sid}}=0$ 
(remember 
that we had already fixed $q_L=0$, as a simplifying assumption). 
They are displayed for three values of the average pair 
momentum, $K_T = 0.17, 0.47$, and $0.80$ GeV/c, and emissivity $\kappa=1$. 

	In Fig. 5, we illustrate the behavior of the correlation functions,  
$C(q_{T_{out}})$ vs. $q_{T_{out}}$ (solid curves) and $C(q_{T_{sid}})$ vs. 
$q_{T_{sid}}$ 
(dashed curves), similarly to Fig.4, but with emissivity $\kappa=0.5$. 
Again, the curves were estimated for $K_T = 0.17, 0.47$, and $0.80$ GeV/c. 
We see that in this case the correlation functions versus $q_{T_{out}}$ are 
always narrower (consequently, the radii are bigger) than those curves 
versus $q_{T_{sid}}$. This originates in the higher contribution from the 
volumetric term with respect to the surface ones, coming from their lower 
emissivity in this case, since $\kappa=0.5$ in that plot. The larger emission 
duration is also a consequence of this reduced surface emissivity with 
respect to the previous case discussed in Fig. 4. 
 
	Although not shown for sake of clarity in the plots, the 
correlation functions, $C(q_{T_{out}})$ x $q_{T_{out}}$ and $C(q_{T_{sid}})$ x 
$q_{T_{sid}}$, 
were also computed for other three values of $K_T$, i.e., all together, 
for $K_T$=0.17, 0.27, 0.38, 0.47, 0.63, and 0.80 GeV/c. Each of the curves, 
either for $\kappa=0.5$ or for $\kappa=1$, were fitted by Gaussian distributions, 
in the regions were their behavior could be reasonably well 
approximated to that distribution. 
In this way, we obtained the corresponding average values of $R_{out}$ 
and $R_{sid}$, as shown in Fig. 6 and 7, 
and from those, we estimated the ratio $R_{out}/R_{sid}$. 
We see that $R_{sid}$ as a function of $K_T$ is basically flat in both cases, 
since we have considered no transverse flow in the computations. However, 
our $R_{out}$ decreases with increasing $K_T$, although not as much as 
suggested by data. This is a consequence of the time
dependence of the temperature, that the higher momentum particles are 
emitted at earlier times.    
The values obtained for the ratio are  plotted in Fig. 8 
together with the preliminary STAR 
(filled triangles) and PHENIX (filled circles) data for both $\pi^+$ and 
$\pi^-$. We can see that our results were highly successful in 
describing both sets of data for $\kappa=1$, but the curve corresponding 
to $\kappa=0.5$ is away above the data limits, suggesting that we should 
have a high emissivity along the system history in order to explain the 
data trend.

\bigskip

\section{Summary and Conclusions}

This simple model works well for the ratio of $R_{out}/R_{sid}$ and suggests
that the origin of the experimental value lies in the opacity of source 
emission and the relatively short time of decoupling of the longitudinal 
expansion, i.e., of the order of the nuclear radius. 
The principal reason why we are able to get such a small ratio of
$R_{out}/R_{side}$ is probably due to a combination of two effects.
The first is that the surface is opaque, and whatever is emitted from the 
surface will have a small value of this radius.  The second effect 
is that we allow black body radiation by gluons when the surface
is very hot.  This allows a much larger contribution from surface 
emission than is typical of what happens in hydrodynamical simulations where
particles are emitted by Cooper-Frye decoupling from a surface at very low
temperature.  In fact we find that about $80\%$ of the emission comes
from the surface.  The fact that so many particle are emitted from the 
surface at early times also means that the longitudinal decoupling time
in this computation is significantly 
shorter than would be the case for hydrodynamic
simulations, and this again goes in the direction suggested by the 
RHIC data, where the longitudinal time scale did not grow as much as might
have been expected between SPS and RHIC energies. 

The model also describes the typical source radii reasonably
well, but not the $K_T$ dependence of these radii, as seen in Figs. 6 and 7.  
This suggests that the time variation of the emitting radius and 
the introduction of transverse flow may play a significant role, 
\cite{heinzkolb}-\cite{teaney}. 
If there is a time variation of the various radii, this will be
correlated with the typical momentum scale of emitted particles, since the
earlier is the time, the hotter are the particles.  We are sensitive
to such variation since we allow emission from the hot surface at early time.
Also, a proper treatment of the decoupling is not included in our 
computations and this certainly will affect the results, although it might 
also suggest modification in the treatment of decoupling, 
\cite{teaney},\cite{bass}.

At a minimum, these computations suggest that the problem in describing
the various HBT ratios lies not so much in $R_{out}/R_{sid}$ as it does
in computing the full set of radii and obtaining a comprehensive and complete
description of all of the above within one dynamical model.  The ratio 
of $R_{out}/R_{sid}$ may very well be independent of many of the variations
within these models since it is dimensionless, where the dimensionful
values of the various radii are sensitive to changes of time and size scales.
Such a complete dynamical computation allowing the possibility of surface
emission at very early times has not been implemented.  It may in fact be
more complicated than we suggest:  Perhaps the surface emissivity is
a strong function of the momentum of the emitted particles.  We certainly 
expect that high momentum particles are more easily emitted than are low 
momentum ones.

\vfill\eject
\acknowledgments
S.S.P. would like to express her gratitude to  
the Nuclear Theory Group at BNL for their kind hospitality 
and for the stimulating discussion atmosphere 
during the elaboration of this work. She would also like to thank 
R. Keith Ellis and the Theoretical Physics Dept. at Fermilab, 
for their kind hospitality. Both of the authors would like to thank Ulrich 
Heinz and Peter Kolb for discussions on an earlier version of this manuscript, 
and the referees of the paper for their meticulous reading and suggestions, 
pointing out an error in the estimate of the multiplicity.
This research was partially supported by CNPq under Proc. N. 200410/82-2. 
This manuscript has been authored under Contracts No. DE-AC02-98CH10886 
and No. DE-AC02-76CH0300 with the U.S. Department of Energy.

\vfill\eject
\begin{figure}
\begin{center}
\epsfig{file=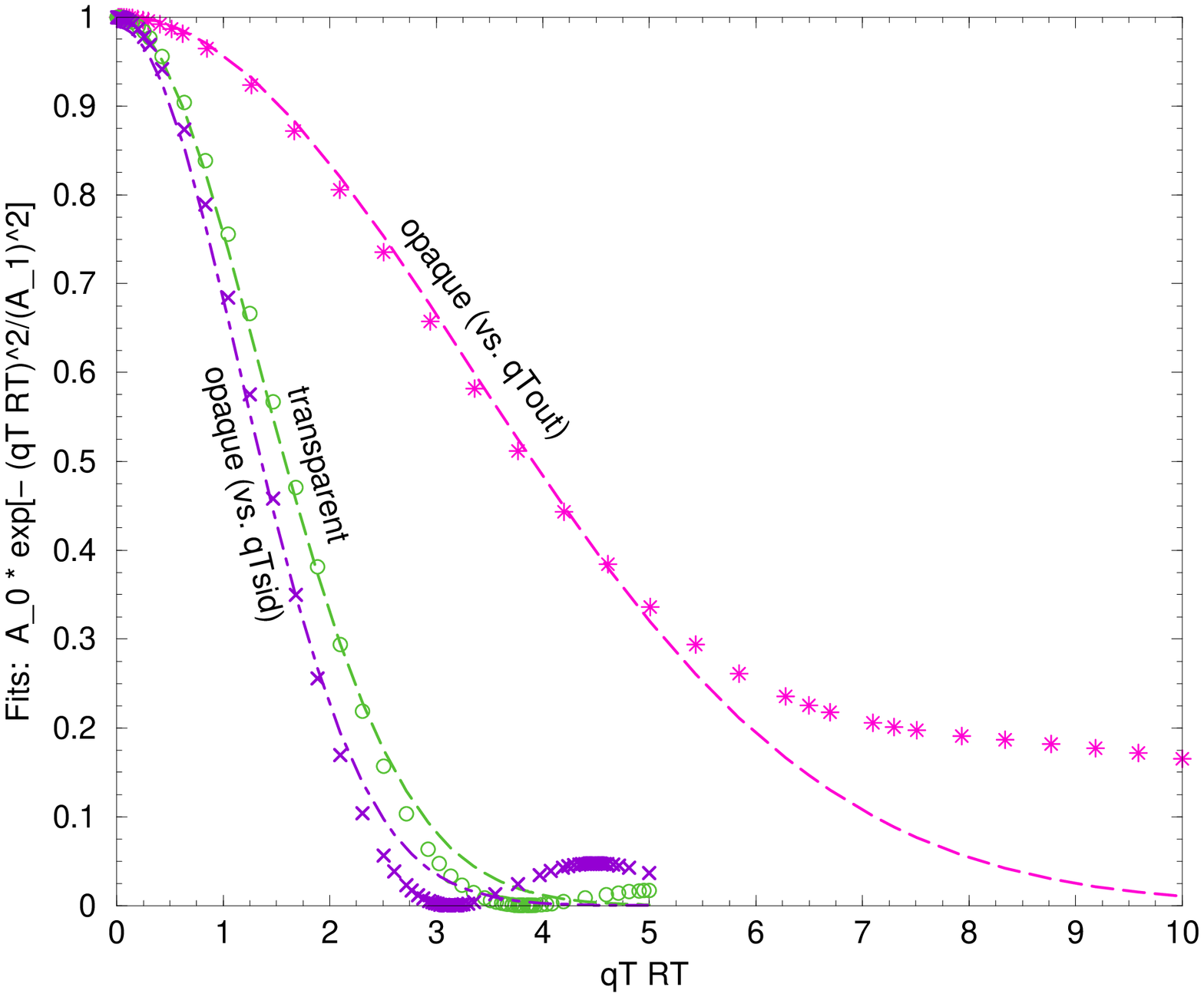,height=11cm,angle=-90}
\vspace{1.5cm}
\end{center}
\caption{
Illustration of the correlation functions in the {\sl very simple model for 
$R_{out}$ and $R_{sid}$} as a function of the corresponding variable 
$q_{T_{out}}$ and $q_{T_{sid}}$. 
For emphasizing the geometrical differences, we considered 
$\Delta t=0$ in the plots of the C($q_{T_{out}}$) vs. $q_{T_{out}}$. 
The set of points and the fitting curve in the middle correspond to both 
C($q_{T_{out}}$) vs. $q_{T_{out}}$ and C($q_{T_{sid}}$) vs. $q_{T_{sid}}$ in 
the transparent case, since {\bf no} time dependence is included 
in the $q_{T_{out}}$ variable in 
the above plot. The narrower and the wider sets correspond, respectively, 
to C($q_{T_{sid}}$) vs. $q_{T_{sid}}$ and to C($q_{T_{out}}$) vs. $q_{T_{out}}$ in the opaque 
case. We see that, as a result of the opacity of the source, this last set 
is much broader than the first one. }
\label{simple_example}
\end{figure}

\vfill\eject
\begin{figure}
\begin{center}
\epsfig{file=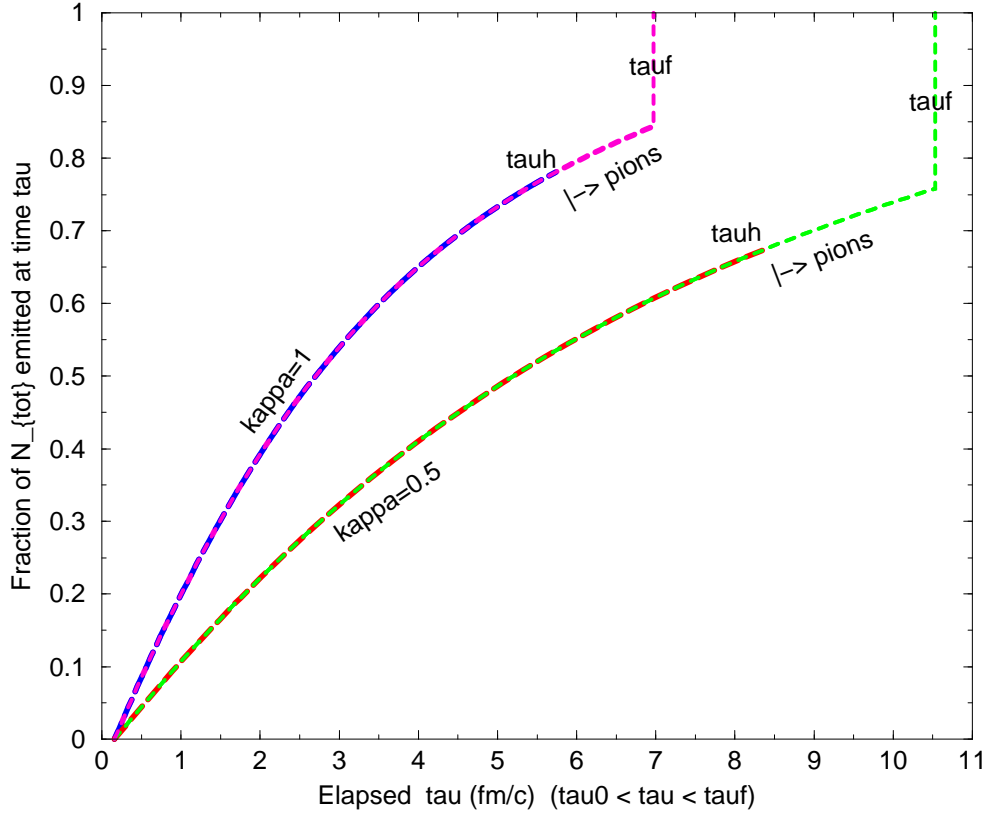,height=11cm,angle=-90}
\vspace{1.5cm}
\end{center}
\caption{
The flux of emitted particles per unit time is shown, normalized to 
their total number, for two values 
of the emissivity ($\kappa=1$ and $\kappa=0.5$), at each instant,  
starting at $\tau_0 = 0.16$ fm/c, passing by the beginning of the 
phase transition, at $\tau=\tau_c$, then through its end at $\tau_h$, 
and finally, stopping at the instant when the pionic system reaches 
$\tau_f$. We clearly see the fraction of the pions still in the 
system at that instant, which is then immediately emitted from the 
entire volume. In this example, we fixed $T_0 = 411$ MeV, $T_c = 175$ MeV, 
$T_f = 150$ MeV, $R_T \approx 7$ fm/c, and the number of quark flavors in 
the QGP phase to be 2. }
\label{flux}
\end{figure}

\vfill\eject
\begin{figure}
\begin{center}
\epsfig{file=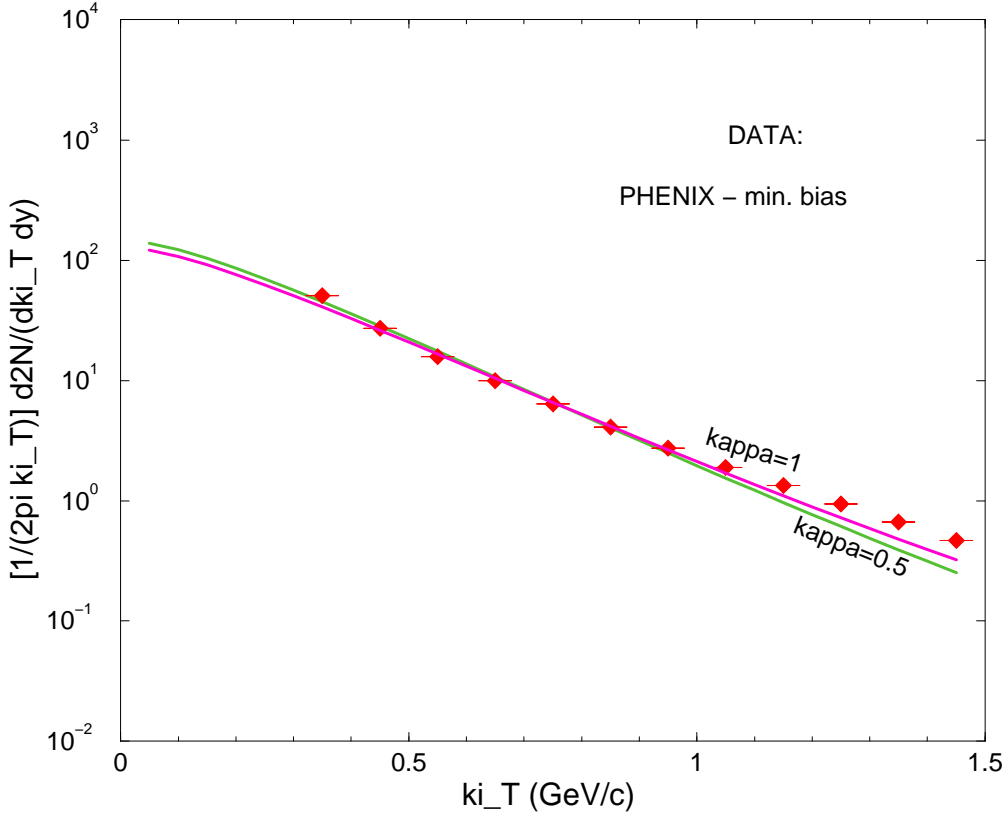,height=11cm,angle=-90}
\vspace{1.5cm}
\end{center}
\caption{
The prediction based on our model for  the transverse momentum 
distribution of emitted pions is shown. The points are from the minimum-bias 
data from PHENIX Collaboration. The curves correspond to 
emissivity $\kappa=0.5$ and to $\kappa=1$, without the inclusion of 
transverse flow. 
We see that both cases describes data on spectrum well 
in the low pion momentum region, up to about  
$k_{i_T}\approx 1$ GeV/c.  The parameters used are explained in the text, 
corresponding to $T_0 = 411$ MeV, $T_c = 175$ MeV, $T_f = 150$ MeV, 
and the transverse radius, $R_T \approx 7$ fm/c. }
\label{f:spectrum}
\end{figure}

\vfill\eject
\begin{figure}
\begin{center}
\epsfig{file=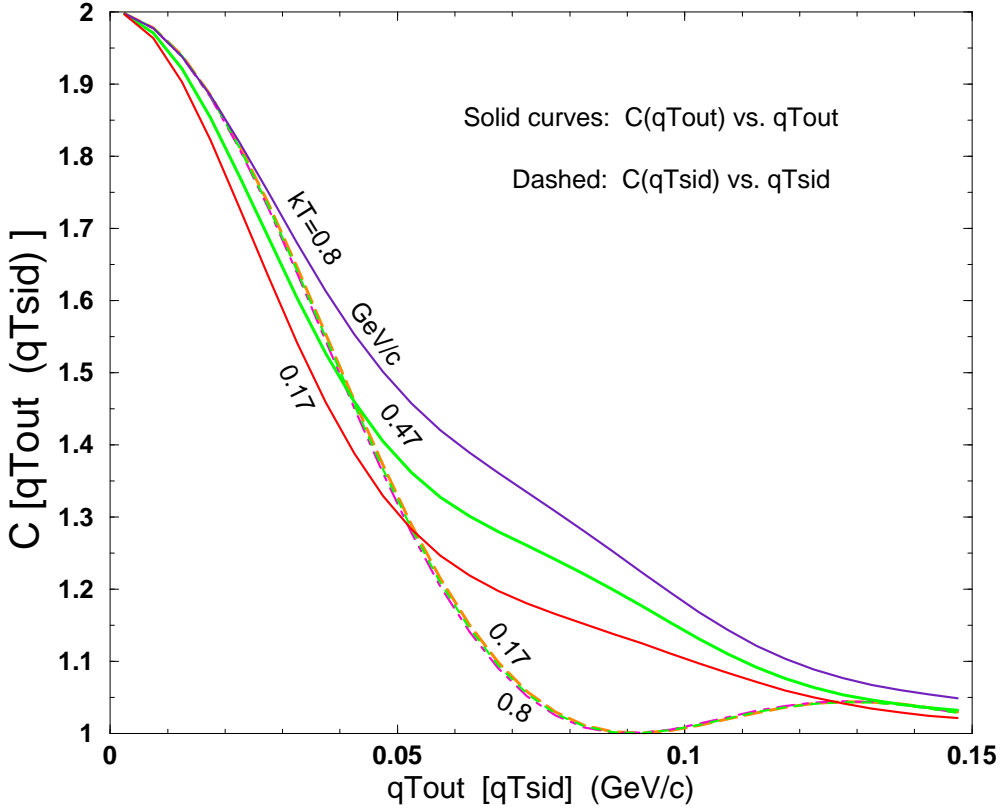,height=11cm,angle=-90}
\vspace{1.5cm}
\end{center}
\caption{
The correlation functions, $C(q_{T_{out}},K_T)$ vs. $q_{T_{out}}$ (solid), and  
$C(q_{T_{sid}},K_T)$ vs. $q_{T_{sid}}$ (dashed), are shown for three 
distinct values of the average pair momentum, $K_T$, corresponding to 
the phase transition temperature $T_c = 175$ MeV, $T_f = 150$ MeV, including 
a volumetric emission when the hadronic (pions) system reaches 
$\tau= \tau_f$. We see that the 
width of the curves as function of $q_{T_{out}}$ 
increase (or conversely, the radii decrease) with 
increasing $K_T$, whereas the curves for different $q_{T_{sid}}$ show no  
visible variation, as would be expected since no transverse 
flow is considered in the computation. The same 
input parameters were adopted here: $T_0 = 411$ MeV, $T_c = 175$ MeV, 
$T_f = 150$ MeV, the transverse radius, 
$R_T \approx 7$ fm/c, and emissivity $\kappa = 1$.}
\label{f:CqtoCqts1}
\end{figure}

\vfill\eject
\begin{figure}
\begin{center}
\epsfig{file=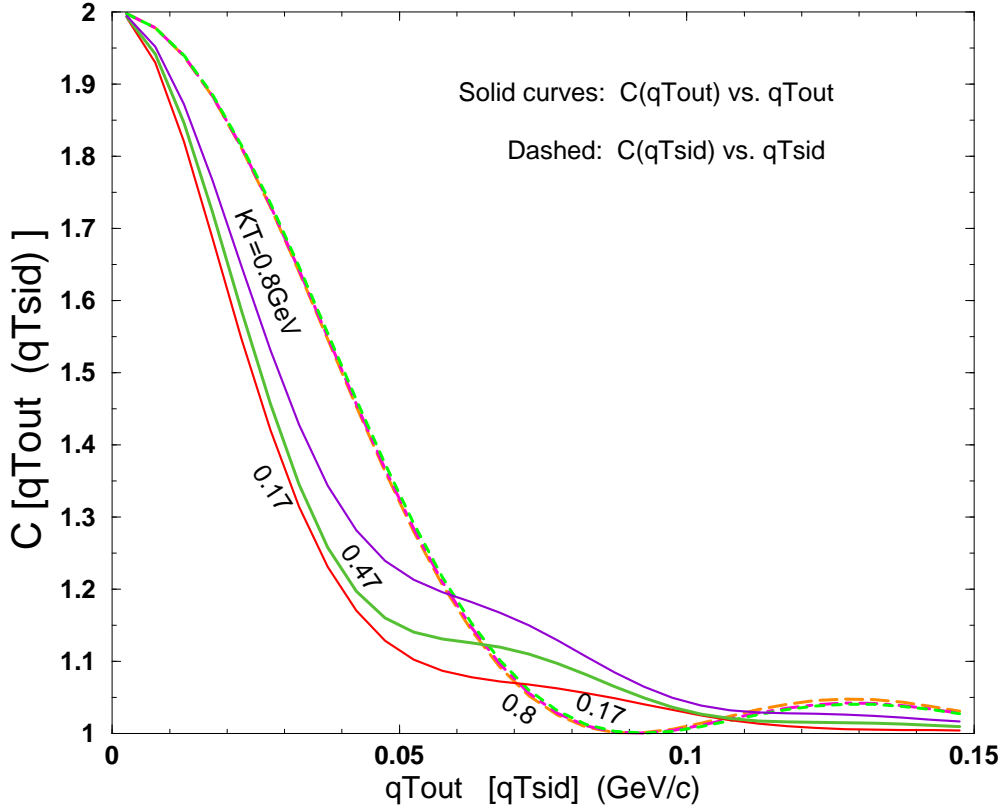,height=11cm,angle=-90}
\vspace{1.5cm}
\end{center}
\caption{
The correlation functions, $C(q_{T_{out}},K_T)$ vs. $q_{T_{out}}$ (solid), and  
$C(q_{T_{sid}},K_T)$ vs. $q_{T_{sid}}$ (dashed), are shown for three 
distinct values of the average pair momentum, $K_T$, as in Fig. 4, but 
now with reduced emissivity, $\kappa =0.5$. 
We see that the width of the solid curves still  
increase (or conversely, the radii decrease) with 
increasing $K_T$, but they are all smaller than the width   
corresponding to the dashed curves. The same 
input parameters as before were adopted here: $T_0 = 411$ MeV, 
$T_c = 175$ MeV, $T_f = 150$ MeV, the transverse radius,  
$R_T \approx 7$ fm/c.}
\label{f:Cqtoqts2}
\end{figure}

\vfill\eject
\begin{figure}
\begin{center}
\epsfig{file=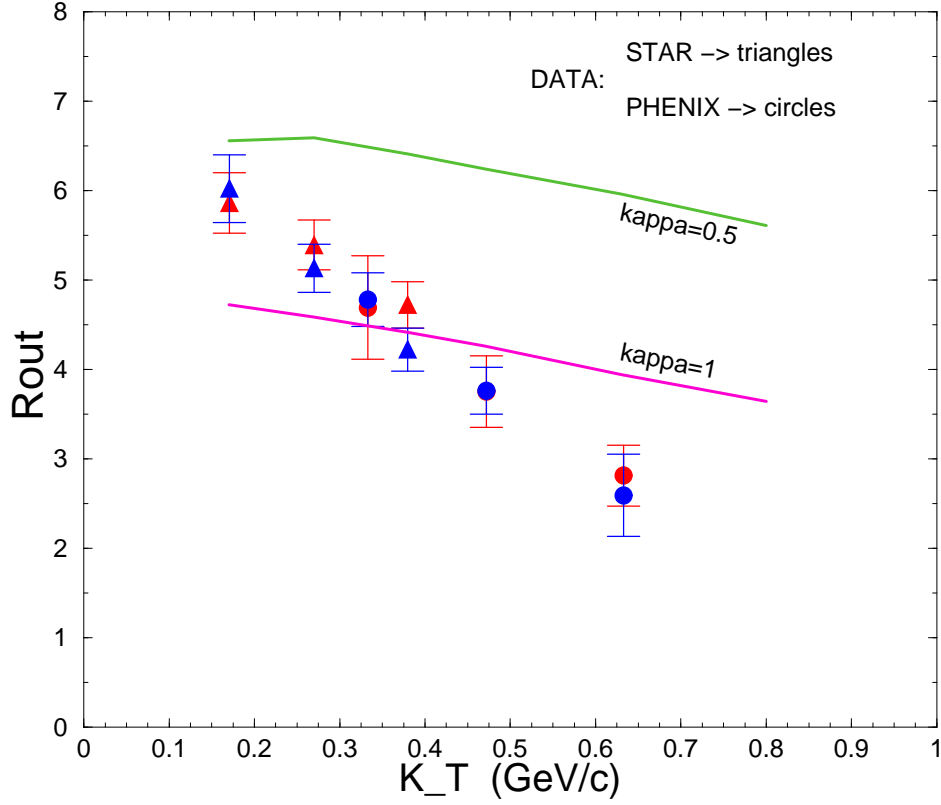,height=11cm,angle=-90}
\vspace{1.5cm}
\end{center}
\caption{
Our results for $R_{out}$ ({\sl outward}) 
radius are shown as a function of the average pair momentum, 
$K_T$, for the two cases discussed before, corresponding to $50\%$ emissivity 
and to $\kappa =1$, without inclusion of transverse flow. The experimental 
data points from STAR (triangles) and PHENIX (circles) are also included in 
the plot. The values of the parameters are 
the same as in the previous plots, i.e., $T_0 = 411$ MeV, $T_c = 175$ MeV, 
$T_f = 150$ MeV, the transverse radius, 
$R_T \approx 7$ fm/c.}
\label{Ro}
\end{figure}

\vfill\eject
\begin{figure}
\begin{center}
\epsfig{file=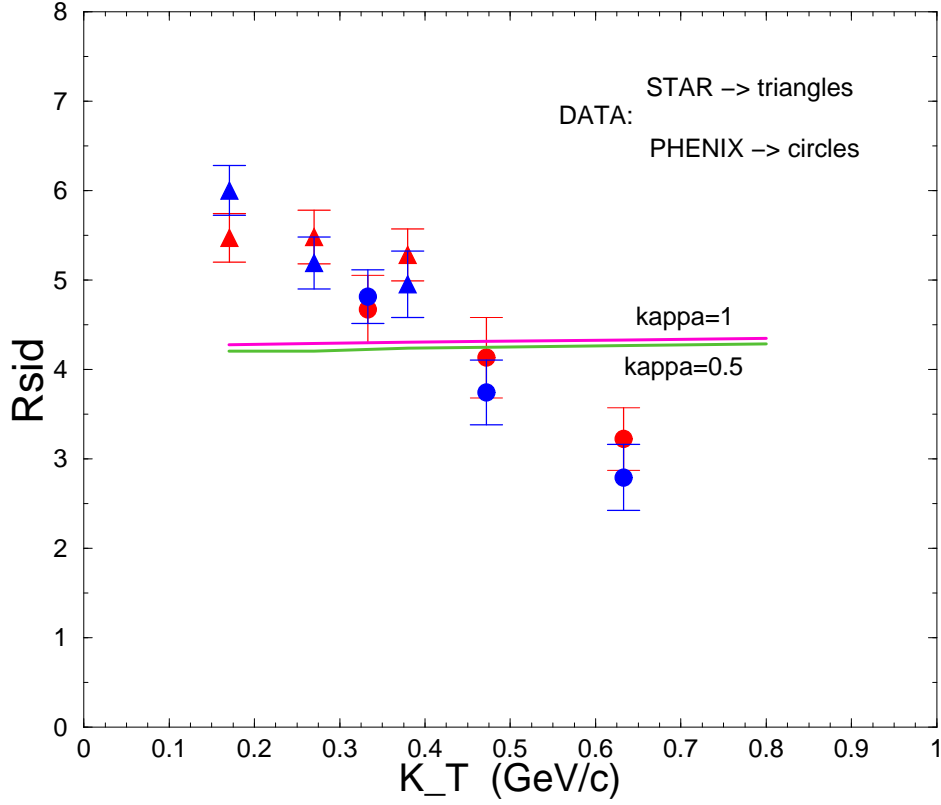,height=11cm,angle=-90}
\vspace{1.5cm}
\end{center}
\caption{
Analogously to the previous case, we show here the results for 
$R_{sid}$ ({\sl sideward}) radius vs. $K_T$, studied for  
$\kappa = 0.5, 1$. No sensitivity to $\kappa$ is seen, since 
no transverse flow is included in the computation. 
The experimental data points with error bars are from 
STAR (triangles) and PHENIX (circles). 
The values of the parameters are 
the same as in the previous plots, i.e., $T_0 = 411$ MeV, $T_c = 175$ MeV, 
$T_f = 150$ MeV, and the transverse radius, $R_T \approx 7$ fm/c.}
\label{Rs}
\end{figure}

\vfill\eject
\begin{figure}
\begin{center}
\epsfig{file=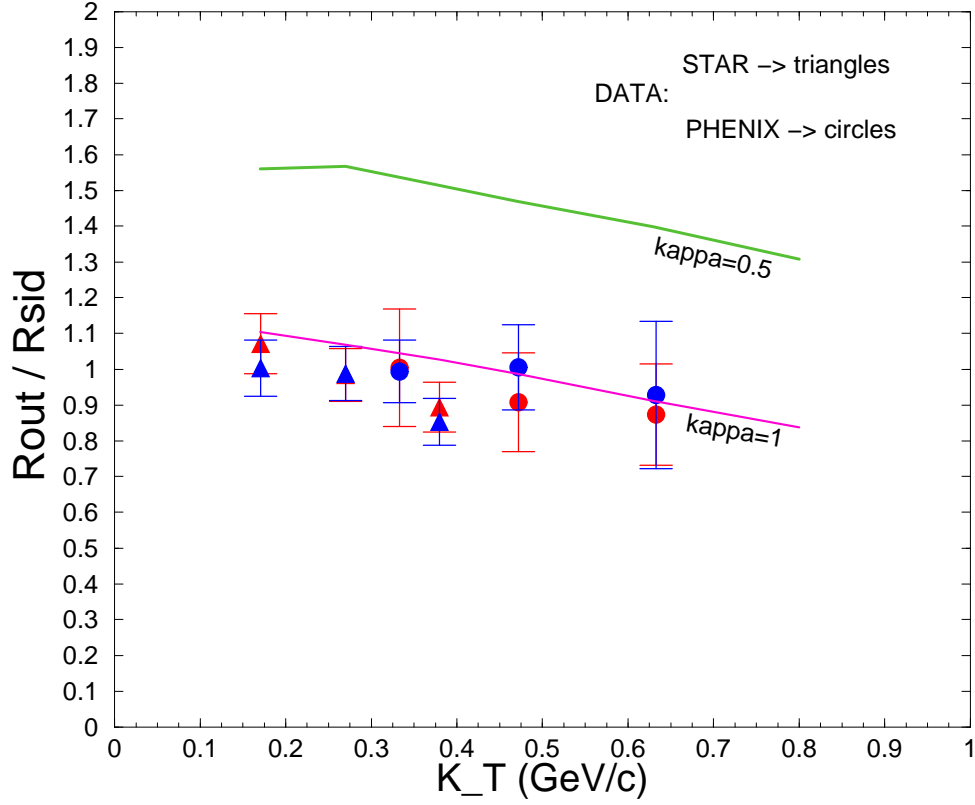,height=11cm,angle=-90}
\vspace{1.5cm}
\end{center}
\caption{
The result corresponding to the ratio $R_{out}/R_{sid}$ of the 
{\sl outward} radius by the  {\sl sideward} one is shown within 
our model. We see that 
the ratio corresponding to full emissivity ($\kappa=1$) agrees very well 
with data within the experimental 
error bars (shown in the plot), whereas the $50\%$ emissivity case is 
completely excluded by data, since 
the ratio is too high in that case, reflecting 
what we saw in the two previous plots. 
The values of the parameters are the same as before, i.e., $T_0 = 411$ MeV, 
$T_c = 175$ MeV, $T_f = 150$ MeV, and the transverse radius, 
$R_T \approx 7$ fm/c.}
\label{f:RoRs1}
\end{figure}

\end{document}